\documentclass[aip,apl,reprint,amsmath,amssymb]{revtex4-2}

\usepackage{graphicx}
\usepackage{xcolor}
\usepackage[colorlinks=true,linkcolor=blue,citecolor=blue,urlcolor=blue]{hyperref}

\begin{document}

\title{Perfectly aperiodic asymmetric scalable metasurface based on the hat monotile}

\author{Yuma Otsuka}

\author{Hiroki Wakatsuchi}
\email[Author to whom correspondence should be addressed: ]{wakatsuchi.hiroki@nitech.ac.jp}
\affiliation{Department of Engineering, Graduate School of Engineering, Nagoya Institute of Technology, Gokiso-cho, Showa, Nagoya, Aichi 466-8555, Japan}

\date{\today}

\begin{abstract}
Metasurfaces are conventionally designed on periodic lattices, which renders their electromagnetic response spatially uniform and leaves the spatial arrangement itself unused as an information-bearing design resource. In this study, we propose and numerically validate a metasurface built on an aperiodic monotile: the recently discovered ``hat'' polykite, a single tile mathematically guaranteed to tile the plane only nonperiodically. The twelve distinguishable orientations of the hat serve as a natural design basis; each carries resonant elements tuned to one of twelve absorption frequencies between 4.600 and 4.914~GHz, transcribing the provable aperiodicity and mirror asymmetry of the geometry into the electromagnetic response. Gaussian-beam scanning over a compositionally unique $6\times 6$ grid yields irregular, position-dependent electromagnetic signatures whose pairwise separations reach beyond 6~dB. In particular, coherence-guided selection of multiple frequencies increases the stable rank of the position--signature matrix from 1.0 to approximately 1.6 with three frequency components, whereas the accumulation of mutually similar electromagnetic features fails to add effective dimensions, demonstrating that properly selected multifrequency characteristics enhance the spatial distinguishability embedded in the surface. A shift-correlation analysis in which the cosine similarity never exceeds 0.5 confirms that the response inherits the absence of translational order from the tiling. Compatible with standard printed-circuit fabrication, the aperiodic metasurface design establishes a deterministic route toward position-identifiable electromagnetic environments for localization, chipless identification, and integrated sensing and communication.
\end{abstract}

\maketitle

Metamaterials and their two-dimensional counterparts, metasurfaces---arrays of subwavelength resonators---provide electromagnetic responses unavailable from natural materials, enabling thin resonant absorbers,\cite{sievenpiper1999,landy2008} negative refraction and superlensing,\cite{shelby2001,pendry2000} cloaking,\cite{schurig2006} and generalized wavefront shaping.\cite{yu2011,pfeiffer2013,li2018,glybovski2016} Digitally coded and reconfigurable implementations extend these capabilities toward programmable surfaces and smart radio environments,\cite{cui2014,direnzo2020} including surfaces whose responses are selected by the temporal waveform of the incident signal.\cite{wakatsuchi2013,ushikoshi2023}

Almost all of these functions, however, are designed on periodic lattices under the implicit assumption of a spatially uniform response. Even wavefront-shaping metasurfaces whose unit cells vary locally may retain periodicity at the supercell level,\cite{yu2011,pfeiffer2013,omori2026experimental} such that the same electromagnetic behavior is repeated across the aperture. In other cases, the design cannot be extended indefinitely and aperiodically, so that the spatial degrees of freedom offered by aperiodicity are confined to a limited area.\cite{yu2011,pfeiffer2013,fathnan2022method} Infinite aperiodicity has been introduced deliberately in one-dimensional impedance profiles synthesized for wide-angle reflection control\cite{budhu2023} and in absorbers arranged according to Fibonacci- or Thue--Morse-type sequences,\cite{yang2017,dalnegro2012} while disorder-engineered metasurfaces have enlarged the accessible design space in optics.\cite{jang2018} Nevertheless, sequence-generated layouts remain tied to their generating axes and typically retain mirror symmetries, one-dimensional syntheses generalize poorly to two dimensions, and random layouts offer no structural guarantee: their lack of order is statistical rather than provable. Consequently, a one-to-one correspondence between the position on a surface and its electromagnetic response---a property that would turn the scattering of a surface into a spatial fingerprint for localization, backscatter identification, physically anchored spatial keys, and emerging integrated sensing and communication (ISAC) networks\cite{hechan2016,wymeersch2020,preradovic2010,delhougne2018,liu2022}---has remained beyond the reach of systematic design.

\begin{figure}[!t]
\includegraphics[width=\columnwidth]{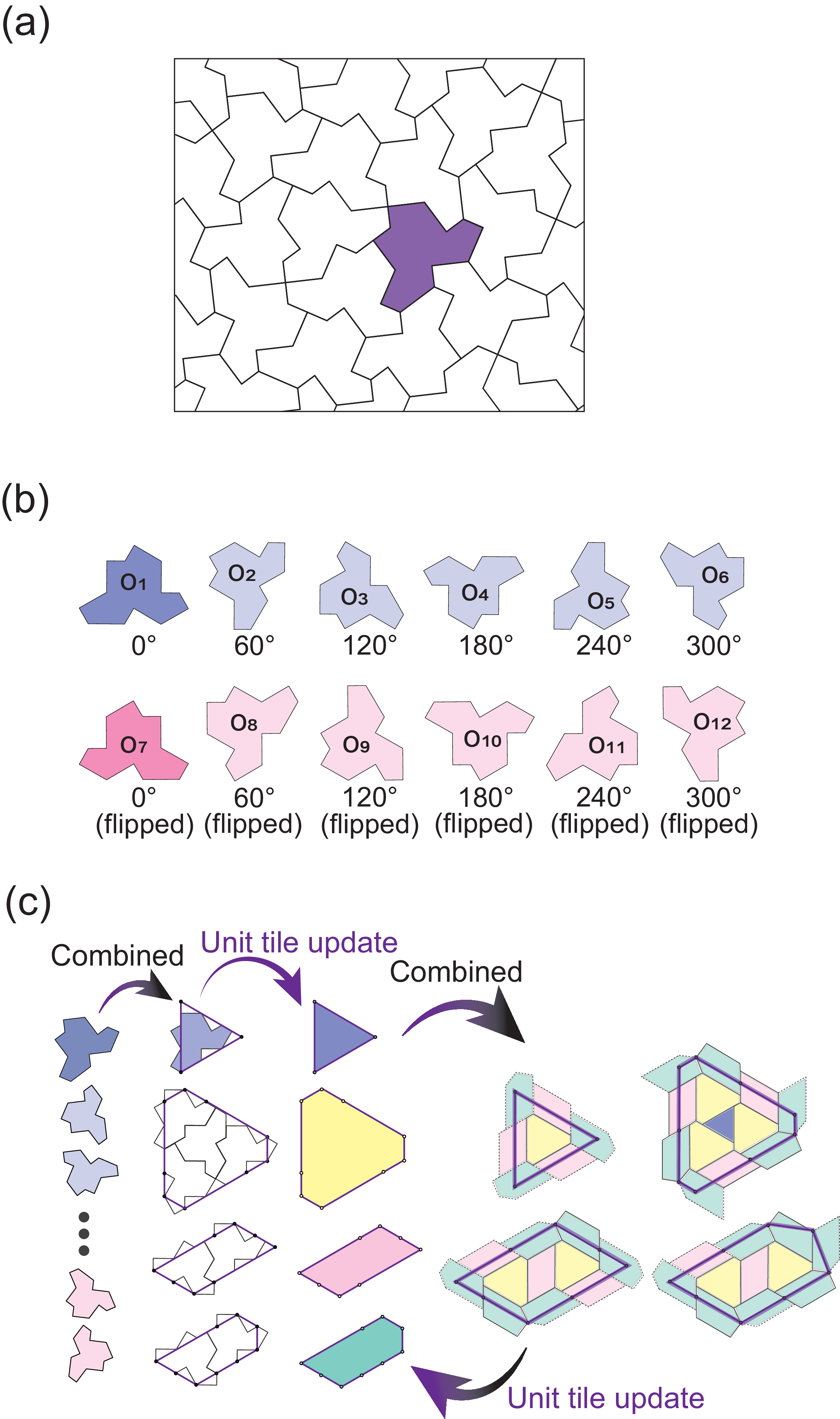}
\caption{\label{fig1} Perfectly aperiodic asymmetric layout based on the hat monotile. (a)~Example of the hat tiling used for the metasurface design. The purple geometry represents a single hat. (b)~The twelve distinguishable hats O$_1$--O$_{12}$ generated from a single hat by $60^\circ$ rotations (top row) and by flipping (bottom row). (c)~Hierarchical construction. Hats are clustered into the four metatiles corresponding to T, H, P, and F tiles (from top to bottom), which are recombined and simplified into updated unit tiles to form successively larger supertiles in an aperiodic manner, guaranteeing unlimited aperiodic expansion.}
\end{figure}

Such a structural guarantee has recently emerged from mathematical research. The ``hat'' polykite was shown to be an aperiodic monotile: a single 13-sided shape that tiles the plane but never periodically (Fig.~\ref{fig1}(a)).\cite{smith2024hat,smith2024spectre} Unlike Penrose-type quasicrystalline patterns, which require at least two tile species and retain high-order rotational and mirror symmetries,\cite{penrose1974,vardeny2013} the hat achieves this guarantee with a single tile that itself possesses no spatial mirror symmetry. Hat tilings are quasiperiodic\cite{socolar2023} and have already been utilized in physics and engineering: monotile lattices host graphene-like spectra with chiral zero modes,\cite{schirmann2024} hat-based quasilattices produce chiral optical diffraction,\cite{moritake2026chiral} the tiling has been applied to element placement in phased arrays,\cite{qiang2025} and, in optical metasurfaces, the orientational diversity of identical scatterers placed on the hat tiling has been exploited to average out angle and polarization sensitivity.\cite{lee2025aperiodic} These studies use the spectral, diffractive, array-thinning, or homogenizing consequences of aperiodicity. The defining combinatorial feature of the hat---every tiling necessarily mixes twelve distinguishable tiles, six $60^\circ$ rotations and six of their mirror images---has, however, not been used as an encoding resource: assigning a distinct electromagnetic property to each of the twelve orientations transcribes the provably nonrepeating arrangement of the tiles into an equally nonrepeating spatial distribution of electromagnetic features. Here, we exploit this property to design what we term a perfectly aperiodic asymmetric metasurface: ``perfectly aperiodic'' because the absence of translational symmetry is guaranteed at all scales by the mathematics of the monotile rather than by statistics, and ``asymmetric'' because the tilings possess no spatial mirror symmetry, in particular none with respect to the two orthogonal axes or the diagonals of the aperture. Each of the twelve oriented hats O$_1$--O$_{12}$ (Fig.~\ref{fig1}(b)) hosts resonant elements tuned to a distinct absorption frequency, so that the guaranteed aperiodicity of the geometry is transcribed into a spatially unique electromagnetic response. By scanning a focused Gaussian beam over the surface, this study shows that (i) the scattering signatures are position-dependent and, at each frequency, spread over several decibels across the aperture, (ii) coherence-guided frequency selection, assessed through the stable rank, enhances the spatial distinguishability of the electromagnetic characteristics, and (iii) the response inherits the absence of translational order.

The geometric foundation is summarized in Fig.~\ref{fig1}(c). Following Smith \emph{et al.},\cite{smith2024hat} clusters of one to four hats form four metatiles named T, H, P, and F tiles whose edges carry matching symbols while maintaining aperiodicity. Combining metatiles according to the substitution rules and replacing each combination with a simplified unit tile yields level-1 supertiles of the same four species. Iterating this ``unit tile update'' (as level-2 supertiles, level-3 supertiles, etc) expands the patch indefinitely while ensuring nonperiodic hat tiling. The substitution is iterated until the patch covers the required aperture.

\begin{figure}[!t]
\includegraphics[width=\columnwidth]{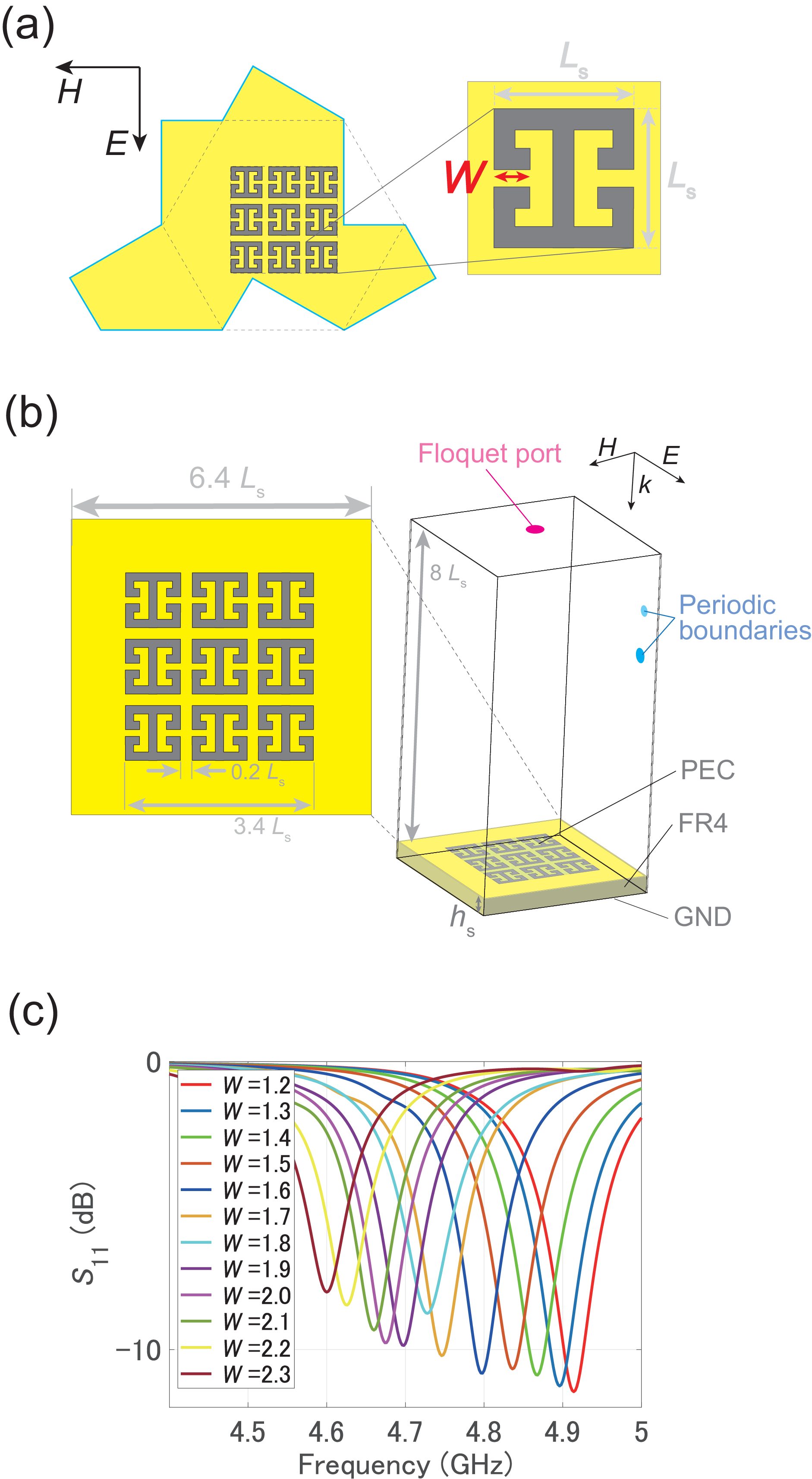}
\caption{\label{fig2}Resonator design. (a)~Conducting resonator elements embedded in each hat. (b)~Simplified model used for the element design. $E$, $H$, and $k$ represent the incident electric field, magnetic field, and wave vector, respectively. (c)~Simulated reflection coefficients for $W=1.2$--2.3~mm. These $W$ values provide twelve distinct absorption resonances between 4.600 and 4.914~GHz, as listed in Table~\ref{tab1}.}
\end{figure}

The spatially unique electromagnetic response is produced by the resonators shown in Fig.~\ref{fig2}(a). Here, each hat hosts a $3\times 3$ array of identical elements. The conducting element (perfect electric conductor) possesses two orthogonal mirror planes but no fourfold rotational symmetry, and all results in this Letter refer to a single linear polarization. Their conducting geometries are designed to resonate internally with weak coupling to their neighbors so that the resonance is determined by the internal elements themselves. This design is important here because otherwise neighboring elements in the aperiodic layout would interfere with the designed resonant mechanism and produce spatially homogenized electromagnetic profiles. The pitch $L_\mathrm{s}=6$~mm is approximately one tenth of the operating wavelength, and the $3\times 3$ grouping averages the local response to stabilize the beam evaluation introduced below. Because the aperiodic layout has no periodic boundaries, the element was designed in the simplified square model of Fig.~\ref{fig2}(b), mimicking the $3\times 3$ array region of a single hat on grounded FR4 (relative permittivity $\varepsilon_\mathrm{r}=4.4$, loss tangent $\tan\delta=0.02$, and thickness $h_\mathrm{s}=1.57$~mm). Sweeping the characteristic width $W$ from 1.2 to 2.3~mm in 0.1-mm steps shifts the reflection minimum from 4.914 to 4.600~GHz (Fig.~\ref{fig2}(c)): widening the metal arms increases the effective capacitance, lowering the $LC$ resonant frequency at which incident power is dissipated in the lossy substrate. The resulting twelve values of $W$ are assigned to hats O$_1$--O$_{12}$, as listed in Table~\ref{tab1}. Here, the resonant frequencies are assigned alternately to the unreflected hats (O$_1$--O$_6$) and the reflected (flipped) hats (O$_7$--O$_{12}$). Thus, the more abundant unreflected hats are separated by larger frequency steps, resulting in a surface in which orientation---and therefore, through tiling, position---is encoded spectrally.

\begin{table}[tb]
\caption{\label{tab1}Design parameter $W$ and simulated resonance frequency assigned to each of the twelve hats (Fig.~\ref{fig1}(b)).}
\begin{ruledtabular}
\begin{tabular}{ccc}
Hat & $W$ (mm) & Resonant frequency (GHz) \\
\hline
O$_1$    & 1.2 & 4.914 \\
O$_2$    & 1.4 & 4.868 \\
O$_3$    & 1.6 & 4.798 \\
O$_4$    & 1.8 & 4.728 \\
O$_5$    & 2.0 & 4.674 \\
O$_6$    & 2.2 & 4.626 \\
O$_7$    & 1.3 & 4.896 \\
O$_8$    & 1.5 & 4.836 \\
O$_9$    & 1.7 & 4.746 \\
O$_{10}$ & 1.9 & 4.698 \\
O$_{11}$ & 2.1 & 4.660 \\
O$_{12}$ & 2.3 & 4.600 \\
\end{tabular}
\end{ruledtabular}
\end{table}

\begin{figure}[!t]
\includegraphics[width=0.82\columnwidth]{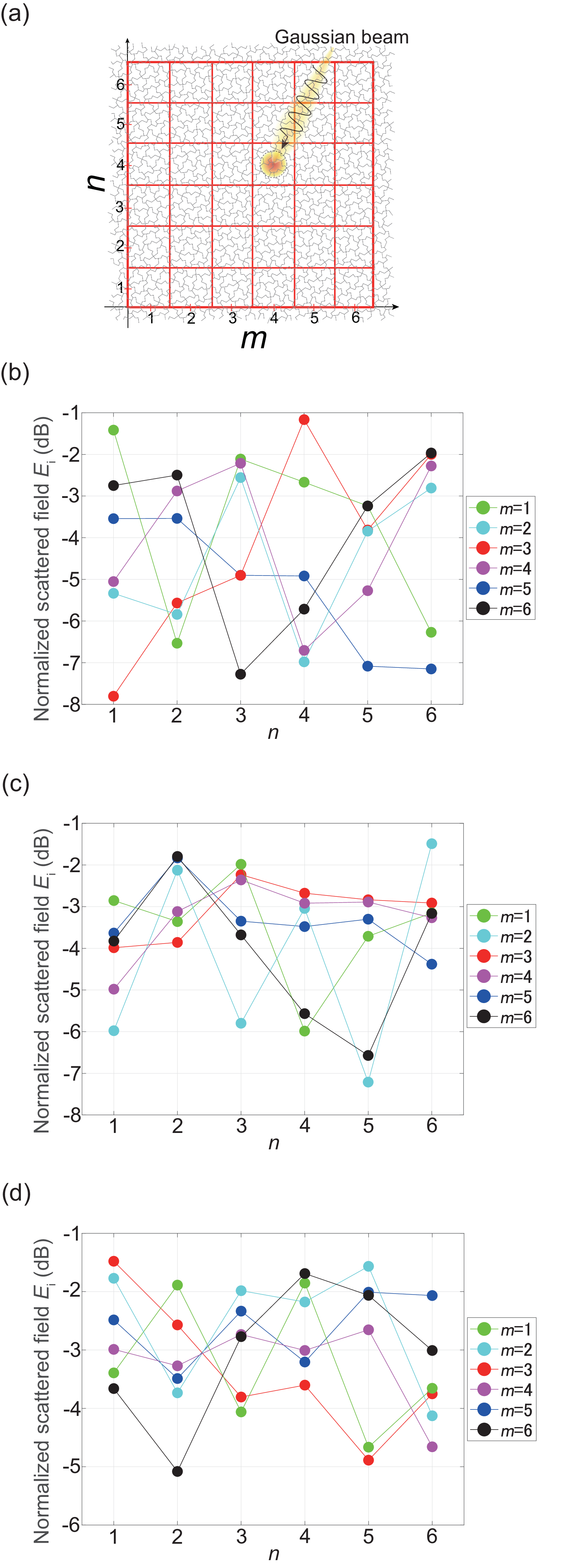}
\caption{\label{fig3}Position-dependent scattering. (a)~Simulation setup: the surface is divided into a $6\times 6$ grid of 115-mm cells indexed by $(m,n)$, and a Gaussian beam is focused onto the center of each cell. (b)--(d)~Normalized scattered field $E_i$ at (b)~$f_\mathrm{a}=4.896$~GHz, (c)~$f_\mathrm{b}=4.798$~GHz, and (d)~$f_\mathrm{c}=4.626$~GHz, plotted along $n$ for each $m$. These three frequencies are those selected by the coherence-guided procedure of Fig.~\ref{fig4}; the remaining nine frequencies are not shown.}
\end{figure}

The full aperiodic surface was then evaluated with the finite-element solver ANSYS HFSS (Electronics Desktop 2023~R2). As shown in Fig.~\ref{fig3}(a), a $690\times 690$~mm$^2$ surface is prepared and divided into a $6\times 6$ grid of 115-mm cells, and a Gaussian beam at normal incidence, polarized along $n$, is focused onto the center of each cell, whose position is labeled by the integers $(m,n)$. The beam width (70~mm in diameter, i.e., a $1/e$ field radius of 35~mm) is smaller than the grid pitch, so that the incident field at the edges of the target grid cell is approximately 24~dB lower than its on-axis value, and the pitch exceeds the free-space wavelength (61--65~mm) at all twelve frequencies. With this pitch, counting the numbers of O$_1$--O$_{12}$ contained in each cell confirms that all 36 cells possess mutually distinct compositions of the twelve orientations. To keep the computations tractable, each illumination was analyzed in a truncated subdomain of $200\times 200$~mm$^2$ centered on the target cell and terminated by radiation boundaries; the influence of the truncation is limited because the incident power at the subdomain edge is at least 60~dB lower than its on-axis value and the elements couple only weakly. For each illumination, the scattered far-field electric field in the surface-normal (specular) direction was recorded at the twelve resonance frequencies $f_k$ ($k=1,\dots,12$; Table~\ref{tab1}) and normalized to that of a perfectly conducting plate of the same size under identical illumination. The normalized value in decibels, $E_i(f_k)$, is referred to as the normalized scattered field of grid cell $i=1,\dots,36$, corresponding to $(m,n)=(1,1),(1,2),\dots,(6,6)$; the $36\times 12$ values form the position--signature matrix. All the 3$\times$3 arrays are deployed at the center of hats, while their orientations are adjusted to maximize the resonant mechanisms of the full aperiodic surface. In other words, central conductor bars are ensured to be in parallel with the incident electric field as shown in Fig.\ \ref{fig2}(a) and Fig.\ \ref{fig2}(b). Figs.~\ref{fig3}(b)--\ref{fig3}(d) display $E_i$ at the three frequencies selected below: the response varies irregularly by approximately 4--7~dB across the aperture, with no visible pattern along $m$, $n$, or the diagonals and with profiles that differ between frequencies, as expected from the position-dependent mixture of resonator elements.

\begin{figure}[!t]
\includegraphics[width=\columnwidth]{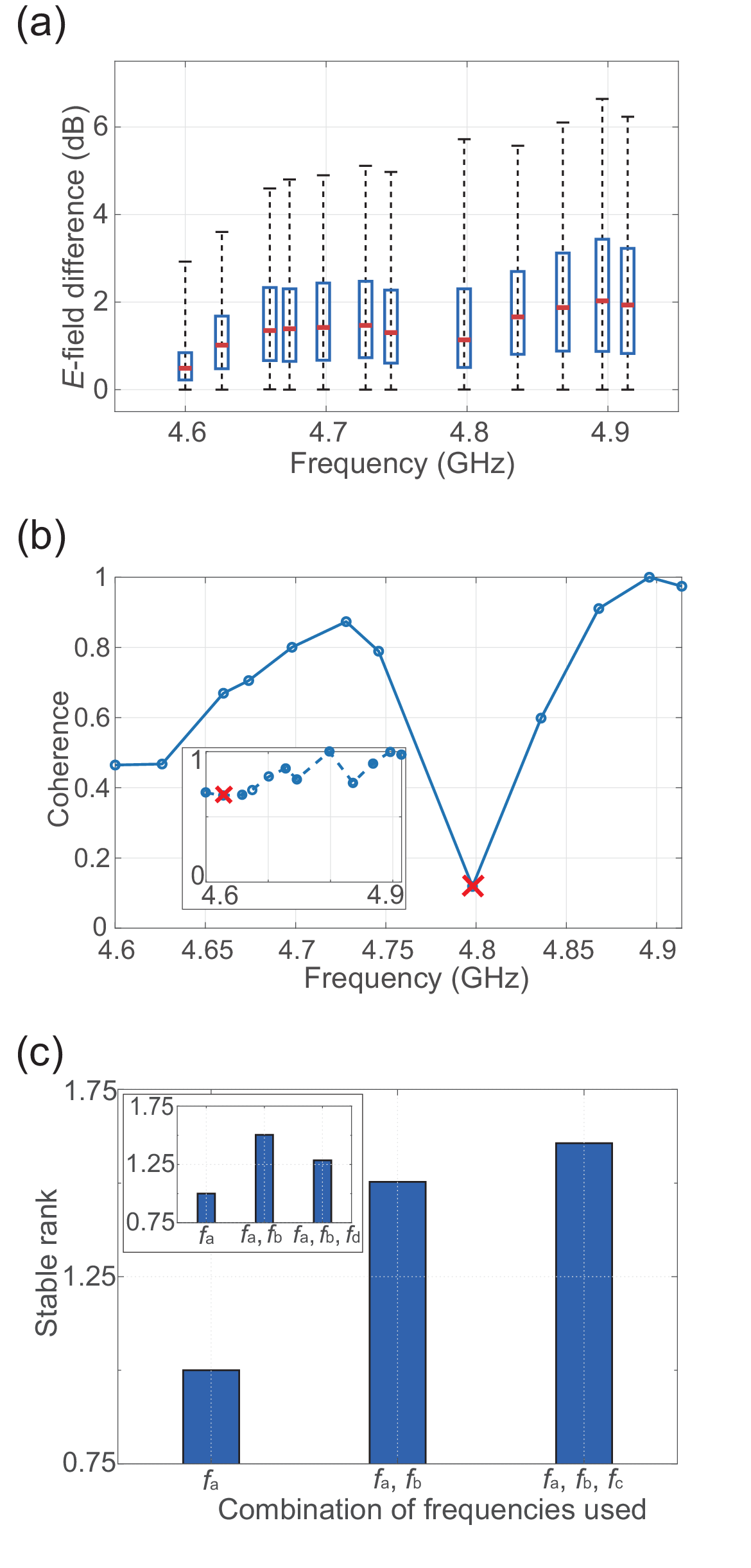}
\caption{\label{fig4}Statistical evaluation of the position signatures. (a)~Box plots of the $E$-field difference $d_{ij}(f_k)$ over all 630 grid-cell pairs at the twelve resonance frequencies. (b)~Coherence $\mu(f_\mathrm{a},f_k)$ with respect to $f_\mathrm{a}=4.896$~GHz; the minimum (cross) occurs at $f_\mathrm{b}=4.798$~GHz. Inset: maximum coherence with respect to $\{f_\mathrm{a},f_\mathrm{b}\}$; its minimum (cross) occurs at $f_\mathrm{c}=4.626$~GHz. (c)~Stable rank $R$ as the least coherent frequencies $f_\mathrm{a}$, $f_\mathrm{b}$, and $f_\mathrm{c}$ are accumulated. Inset: the most coherent remaining frequency $f_\mathrm{d}=4.914$~GHz is added instead of $f_\mathrm{c}$, which reduces $R$.}
\end{figure}

We next quantify how effectively these responses distinguish positions (Fig.~\ref{fig4}). For each frequency, the $E$-field difference $d_{ij}(f_k)=\lvert E_i(f_k)-E_j(f_k)\rvert$ was evaluated over all 630 grid-cell pairs $(=\binom{36}{2})$ (Fig.~\ref{fig4}(a)). Every frequency yields a finite spread, with median separations of 0.5--2~dB and extremes beyond 6~dB, increasing toward the upper band edge. Moreover, the whiskers extend to zero; i.e., at any single frequency, some pairs of cells are nearly indistinguishable, so a single-frequency spatial profile cannot guarantee unique identification. A simple solution is to combine the information of several frequencies, which is effective only if their spatial response patterns are mutually independent. To demonstrate this idea, we form for each frequency the spatial response vector $\mathbf{v}(f_k)=[E_1(f_k),\dots,E_M(f_k)]^{\mathrm T}$, remove its mean via $\tilde{\mathbf{v}}(f_k)=\mathbf{v}(f_k)-\bar v(f_k)\mathbf{1}$ with $\bar v(f_k)=M^{-1}\sum_{i=1}^{M}E_i(f_k)$ and $M=36$ (the number of grid cells), and evaluate the coherence
\begin{equation}
\mu(f_\mathrm{a},f_k)=\frac{\lvert\langle \tilde{\mathbf{v}}(f_\mathrm{a}),\tilde{\mathbf{v}}(f_k)\rangle\rvert}{\lVert\tilde{\mathbf{v}}(f_\mathrm{a})\rVert_2\,\lVert\tilde{\mathbf{v}}(f_k)\rVert_2}.
\label{eq:coherence}
\end{equation}
Here, the reference $f_\mathrm{a}=f_7=4.896$~GHz is chosen because this frequency exhibits the deepest reflection minimum in Fig.~\ref{fig2}(c) and, correspondingly, the largest spread of $d_{ij}$ in Fig.~\ref{fig4}(a), i.e., the strongest spatial contrast. Analogous to compressed sensing, where low mutual coherence between measurement vectors underpins faithful signal recovery,\cite{donoho2006,candes2008} a small $\mu$ indicates that $f_k$ probes a spatial pattern nearly independent of that at $f_\mathrm{a}$. As shown in Fig.~\ref{fig4}(b), the coherence reaches a pronounced minimum of $\mu\approx 0.12$ at $f_\mathrm{b}=f_3=4.798$~GHz, which is selected as the second frequency. In the second round, each remaining candidate is evaluated by its maximum coherence with respect to the frequencies already selected, $\mu_{\max}(f_k)=\max\{\mu(f_\mathrm{a},f_k),\mu(f_\mathrm{b},f_k)\}$ (inset of Fig.~\ref{fig4}(b)); its minimum ($\mu_{\max}\approx 0.66$) identifies $f_\mathrm{c}=f_{6}=4.626$~GHz as the third frequency. These are the frequencies displayed in Figs.~\ref{fig3}(b)--\ref{fig3}(d).

The benefit of using these frequencies can be explained by the stable rank of the accumulated position--signature matrix $X=[\tilde{\mathbf{v}}(f_\mathrm{a}),\tilde{\mathbf{v}}(f_\mathrm{b}),\dots]$. With $\lambda_l$ denoting the eigenvalues of the Gram matrix $G=X^{\mathrm T}X$ and $\lambda_{\max}$ the largest of them,
\begin{equation}
R=\frac{\lVert X\rVert_\mathrm{F}^2}{\lVert X\rVert_2^2}=\frac{\sum_l \lambda_l}{\lambda_{\max}}
\label{eq:srank}
\end{equation}
measures the effective number of independent spatial patterns contained in $X$.\cite{vershynin2018} As shown in Fig.~\ref{fig4}(c), accumulating the least coherent frequencies increases $R$ from 1 to approximately 1.5 ($f_\mathrm{a},f_\mathrm{b}$) and 1.6 ($f_\mathrm{a},f_\mathrm{b},f_\mathrm{c}$); since $R$ is bounded above by the number of accumulated frequencies, this increase indicates that each added frequency contributes independent positional information. In contrast, when the most coherent remaining frequency, $f_\mathrm{d}=f_1=4.914$~GHz ($\mu_{\max}\approx 0.97$), is added as the third frequency instead of $f_\mathrm{c}$, $R$ decreases from 1.5 to approximately 1.3 (inset of Fig.~\ref{fig4}(c)): a nearly parallel column enlarges $\lambda_{\max}$ without adding an independent dimension. Multifrequency assessment therefore amplifies the encoded spatial information, provided that the spatial profiles are mutually independent, with Eq.~(\ref{eq:coherence}) as a simple data-driven selection rule.

\begin{figure}[!t]
\includegraphics[width=0.9\columnwidth]{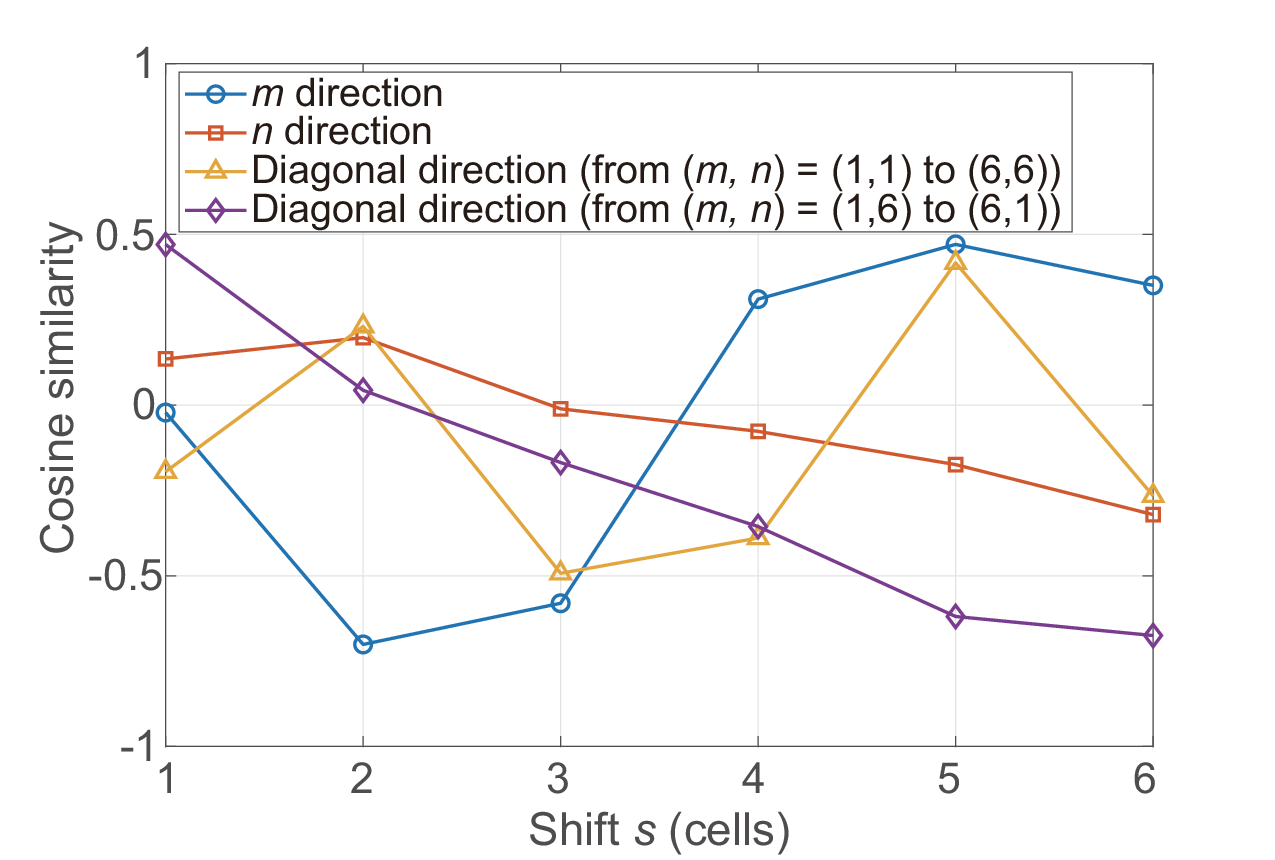}
\caption{\label{fig5}Mean-subtracted cosine similarity $\mathrm{CS}(s)$ between the six-cell response pattern inside the window and the pattern shifted by $s$ cells along a twelve-cell scan, evaluated at $f_\mathrm{a}=4.896$~GHz along the $m$ axis ($n=1$), the $n$ axis ($m=1$), and the two diagonals.}
\end{figure}

Finally, we verify that the electromagnetic response itself---not merely the geometry---is free of translational order (Fig.~\ref{fig5}). At $f_\mathrm{a}$, the beam scan was extended to sequences of twelve consecutive cells along the $m$ axis (row $n=1$), along the $n$ axis (column $m=1$), and along the two diagonals through the window (from $(m,n)=(1,1)$ toward $(6,6)$ and from $(1,6)$ toward $(6,1)$); since the $6\times 6$ window contains only six cells per line, the sequences continue over the same tiling beyond the window. Each twelve-cell sequence $a_\ell$ ($\ell=1,\dots,12$) was first centered by removing its mean over the twelve cells, and the response pattern over the six cells inside the window, $\tilde{\mathbf{x}}^{(0)}=(\tilde{a}_1,\dots,\tilde{a}_6)$, was then compared with the pattern over six consecutive cells shifted by $s$ cells, $\tilde{\mathbf{x}}^{(s)}=(\tilde{a}_{1+s},\dots,\tilde{a}_{6+s})$ with $s=1,\dots,6$, through the mean-subtracted cosine similarity
\begin{align}
\mathrm{CS}(s)&=\frac{\langle \tilde{\mathbf{x}}^{(0)},\tilde{\mathbf{x}}^{(s)}\rangle}{\lVert\tilde{\mathbf{x}}^{(0)}\rVert_2\,\lVert\tilde{\mathbf{x}}^{(s)}\rVert_2},\label{eq:cosine}\\
\tilde{a}_\ell &= a_\ell-\frac{1}{12}\sum_{\ell'=1}^{12} a_{\ell'}, 
\end{align}
where $\ell=1,\dots,12$. For any layout that repeats with a period of up to six grid cells---including supercell-based designs---$\mathrm{CS}(s)$ would return to unity within this range. Instead, as shown in Fig.~\ref{fig5}, the similarity fluctuates between approximately $-0.7$ and $+0.5$ for all four directions and shifts $s=1$--6, never approaching unity. The scattering response thus inherits, within the evaluated window, the aperiodicity that the hat tiling guarantees for the geometry at all scales.

Three features deserve emphasis. First, unlike disorder-based approaches\cite{jang2018} and unlike hat-tiling metasurfaces that use orientational diversity to homogenize the response,\cite{lee2025aperiodic} the aperiodicity here is deterministic, constructive, and used to differentiate rather than to average: the substitution rules generate the layout reproducibly and remain valid for arbitrarily large apertures, so the design scales without reoptimization or statistical verification. Second, the tiling hierarchy imposes a natural matching condition on the assessment: a footprint much smaller than a metatile revisits identical substructures, whereas one spanning many hats averages the composition toward uniformity; the compositional analysis above identifies the intermediate scale at which every grid cell is unique. Third, the intrinsic chirality of the hat---reflected tiles occur far less often than unreflected tiles---invites polarization-sensitive extensions with enantiomer-specific anisotropic elements, and the strictly chiral Spectre monotile\cite{smith2024spectre} offers a reflection-free alternative basis. The position-unique signatures point toward passive, chipless identification in backscatter and RFID-like systems,\cite{preradovic2010} surface-assisted fingerprint localization,\cite{hechan2016,wymeersch2020,delhougne2018} and physically anchored spatial keys for secure wireless links. The scope of the validation is stated explicitly: all the results refer to normal incidence, one linear polarization, and the finite $6\times 6$ window---``perfectly aperiodic'' describes the mathematical guarantee of the layout, while the electromagnetic statements are restricted to the evaluated aperture. In practice, identification under noise and oblique incidence, receiver-weighted metrics, and experimental validation on printed-circuit prototypes remain to be addressed. The proposed method enhances the spatial distinguishability of the electromagnetic response, particularly through the use of multiple frequencies. The degrees of freedom available for encoding can be further increased by exploiting other variables and dependences, such as nonlinearity,\cite{li2017nonlinear,kivshar2003nonlinear} time,\cite{wakatsuchi2013,vellucci2019waveform,barbuto2021metasurfaces} and frequency sequence.\cite{takeshita2024frequency}

In summary, we proposed and numerically validated a metasurface whose layout is nonperiodic and asymmetric, obtained by introducing the twelve orientations of the hat monotile with spectrally distinct absorbing resonators. Gaussian-beam scanning over a compositionally unique $6\times 6$ grid produced irregular position-dependent signatures, coherence-guided frequency selection increased the stable rank of the position--signature matrix, and shift-correlation analysis verified the absence of translational order in the response within the evaluated window. By incorporating the monotile's mathematical guarantee into electromagnetic design, the approach removes the uniform-response constraint of conventional metasurfaces and establishes a deterministic route toward position-identifiable electromagnetic environments for localization, chipless identification, and integrated sensing and communication.

\begin{acknowledgments}
This study was supported in part by the Japan Science and Technology Agency (JST) under the AIP Acceleration Research No.\ JPMJCR25U4, the Japan Society for the Promotion of Science (JSPS) KAKENHI Grant No.\ 26H02139, the Ministry of Internal Affairs and Communications (MIC) under the Fundamental Technologies for Sustainable Efficient Radio Wave Use R$\&$D Project (FORWARD) No.\ JPMI250610002, and the Adopting Sustainable Partnerships for Innovative Research Ecosystem (ASPIRE) No.\ JPMJAP2431.
\end{acknowledgments}

\section*{Author Declarations}
\subsection*{Conflict of Interest}
The authors have no conflicts to disclose.

\subsection*{Author Contributions}
\textbf{Yuma Otsuka}: Data curation (lead); Formal analysis (lead); Investigation (lead); Methodology (equal); Software (lead); Visualization (lead); Writing -- original draft (supporting). \textbf{Hiroki Wakatsuchi}: Conceptualization (lead); Funding acquisition (lead); Methodology (equal); Project administration (lead); Supervision (lead); Writing -- original draft (lead); Writing -- review \& editing (lead).

\section*{Data Availability}
The data that support the findings of this study are available from the corresponding author upon reasonable request.


\begin{thebibliography}{41}%
\makeatletter
\providecommand \@ifxundefined [1]{%
 \@ifx{#1\undefined}
}%
\providecommand \@ifnum [1]{%
 \ifnum #1\expandafter \@firstoftwo
 \else \expandafter \@secondoftwo
 \fi
}%
\providecommand \@ifx [1]{%
 \ifx #1\expandafter \@firstoftwo
 \else \expandafter \@secondoftwo
 \fi
}%
\providecommand \natexlab [1]{#1}%
\providecommand \enquote  [1]{``#1''}%
\providecommand \bibnamefont  [1]{#1}%
\providecommand \bibfnamefont [1]{#1}%
\providecommand \citenamefont [1]{#1}%
\providecommand \href@noop [0]{\@secondoftwo}%
\providecommand \href [0]{\begingroup \@sanitize@url \@href}%
\providecommand \@href[1]{\@@startlink{#1}\@@href}%
\providecommand \@@href[1]{\endgroup#1\@@endlink}%
\providecommand \@sanitize@url [0]{\catcode `\\12\catcode `\$12\catcode `\&12\catcode `\#12\catcode `\^12\catcode `\_12\catcode `\%12\relax}%
\providecommand \@@startlink[1]{}%
\providecommand \@@endlink[0]{}%
\providecommand \url  [0]{\begingroup\@sanitize@url \@url }%
\providecommand \@url [1]{\endgroup\@href {#1}{\urlprefix }}%
\providecommand \urlprefix  [0]{URL }%
\providecommand \Eprint [0]{\href }%
\providecommand \doibase [0]{https://doi.org/}%
\providecommand \selectlanguage [0]{\@gobble}%
\providecommand \bibinfo  [0]{\@secondoftwo}%
\providecommand \bibfield  [0]{\@secondoftwo}%
\providecommand \translation [1]{[#1]}%
\providecommand \BibitemOpen [0]{}%
\providecommand \bibitemStop [0]{}%
\providecommand \bibitemNoStop [0]{.\EOS\space}%
\providecommand \EOS [0]{\spacefactor3000\relax}%
\providecommand \BibitemShut  [1]{\csname bibitem#1\endcsname}%
\let\auto@bib@innerbib\@empty
\bibitem [{\citenamefont {Sievenpiper}\ \emph {et~al.}(1999)\citenamefont {Sievenpiper}, \citenamefont {Zhang}, \citenamefont {Broas}, \citenamefont {Alexopolous},\ and\ \citenamefont {Yablonovitch}}]{sievenpiper1999}%
  \BibitemOpen
  \bibfield  {author} {\bibinfo {author} {\bibfnamefont {D.}~\bibnamefont {Sievenpiper}}, \bibinfo {author} {\bibfnamefont {L.}~\bibnamefont {Zhang}}, \bibinfo {author} {\bibfnamefont {R.~F.~J.}\ \bibnamefont {Broas}}, \bibinfo {author} {\bibfnamefont {N.~G.}\ \bibnamefont {Alexopolous}},\ and\ \bibinfo {author} {\bibfnamefont {E.}~\bibnamefont {Yablonovitch}},\ }\bibfield  {title} {\enquote {\bibinfo {title} {High-impedance electromagnetic surfaces with a forbidden frequency band},}\ }\href@noop {} {\bibfield  {journal} {\bibinfo  {journal} {IEEE Trans. Microw. Theory Tech.}\ }\textbf {\bibinfo {volume} {47}},\ \bibinfo {pages} {2059--2074} (\bibinfo {year} {1999})}\BibitemShut {NoStop}%
\bibitem [{\citenamefont {Landy}\ \emph {et~al.}(2008)\citenamefont {Landy}, \citenamefont {Sajuyigbe}, \citenamefont {Mock}, \citenamefont {Smith},\ and\ \citenamefont {Padilla}}]{landy2008}%
  \BibitemOpen
  \bibfield  {author} {\bibinfo {author} {\bibfnamefont {N.~I.}\ \bibnamefont {Landy}}, \bibinfo {author} {\bibfnamefont {S.}~\bibnamefont {Sajuyigbe}}, \bibinfo {author} {\bibfnamefont {J.~J.}\ \bibnamefont {Mock}}, \bibinfo {author} {\bibfnamefont {D.~R.}\ \bibnamefont {Smith}},\ and\ \bibinfo {author} {\bibfnamefont {W.~J.}\ \bibnamefont {Padilla}},\ }\bibfield  {title} {\enquote {\bibinfo {title} {Perfect metamaterial absorber},}\ }\href@noop {} {\bibfield  {journal} {\bibinfo  {journal} {Phys. Rev. Lett.}\ }\textbf {\bibinfo {volume} {100}},\ \bibinfo {pages} {207402} (\bibinfo {year} {2008})}\BibitemShut {NoStop}%
\bibitem [{\citenamefont {Shelby}, \citenamefont {Smith},\ and\ \citenamefont {Schultz}(2001)}]{shelby2001}%
  \BibitemOpen
  \bibfield  {author} {\bibinfo {author} {\bibfnamefont {R.~A.}\ \bibnamefont {Shelby}}, \bibinfo {author} {\bibfnamefont {D.~R.}\ \bibnamefont {Smith}},\ and\ \bibinfo {author} {\bibfnamefont {S.}~\bibnamefont {Schultz}},\ }\bibfield  {title} {\enquote {\bibinfo {title} {Experimental verification of a negative index of refraction},}\ }\href@noop {} {\bibfield  {journal} {\bibinfo  {journal} {Science}\ }\textbf {\bibinfo {volume} {292}},\ \bibinfo {pages} {77--79} (\bibinfo {year} {2001})}\BibitemShut {NoStop}%
\bibitem [{\citenamefont {Pendry}(2000)}]{pendry2000}%
  \BibitemOpen
  \bibfield  {author} {\bibinfo {author} {\bibfnamefont {J.~B.}\ \bibnamefont {Pendry}},\ }\bibfield  {title} {\enquote {\bibinfo {title} {Negative refraction makes a perfect lens},}\ }\href@noop {} {\bibfield  {journal} {\bibinfo  {journal} {Phys. Rev. Lett.}\ }\textbf {\bibinfo {volume} {85}},\ \bibinfo {pages} {3966--3969} (\bibinfo {year} {2000})}\BibitemShut {NoStop}%
\bibitem [{\citenamefont {Schurig}\ \emph {et~al.}(2006)\citenamefont {Schurig}, \citenamefont {Mock}, \citenamefont {Justice}, \citenamefont {Cummer}, \citenamefont {Pendry}, \citenamefont {Starr},\ and\ \citenamefont {Smith}}]{schurig2006}%
  \BibitemOpen
  \bibfield  {author} {\bibinfo {author} {\bibfnamefont {D.}~\bibnamefont {Schurig}}, \bibinfo {author} {\bibfnamefont {J.~J.}\ \bibnamefont {Mock}}, \bibinfo {author} {\bibfnamefont {B.~J.}\ \bibnamefont {Justice}}, \bibinfo {author} {\bibfnamefont {S.~A.}\ \bibnamefont {Cummer}}, \bibinfo {author} {\bibfnamefont {J.~B.}\ \bibnamefont {Pendry}}, \bibinfo {author} {\bibfnamefont {A.~F.}\ \bibnamefont {Starr}},\ and\ \bibinfo {author} {\bibfnamefont {D.~R.}\ \bibnamefont {Smith}},\ }\bibfield  {title} {\enquote {\bibinfo {title} {Metamaterial electromagnetic cloak at microwave frequencies},}\ }\href@noop {} {\bibfield  {journal} {\bibinfo  {journal} {Science}\ }\textbf {\bibinfo {volume} {314}},\ \bibinfo {pages} {977--980} (\bibinfo {year} {2006})}\BibitemShut {NoStop}%
\bibitem [{\citenamefont {Yu}\ \emph {et~al.}(2011)\citenamefont {Yu}, \citenamefont {Genevet}, \citenamefont {Kats}, \citenamefont {Aieta}, \citenamefont {Tetienne}, \citenamefont {Capasso},\ and\ \citenamefont {Gaburro}}]{yu2011}%
  \BibitemOpen
  \bibfield  {author} {\bibinfo {author} {\bibfnamefont {N.}~\bibnamefont {Yu}}, \bibinfo {author} {\bibfnamefont {P.}~\bibnamefont {Genevet}}, \bibinfo {author} {\bibfnamefont {M.~A.}\ \bibnamefont {Kats}}, \bibinfo {author} {\bibfnamefont {F.}~\bibnamefont {Aieta}}, \bibinfo {author} {\bibfnamefont {J.-P.}\ \bibnamefont {Tetienne}}, \bibinfo {author} {\bibfnamefont {F.}~\bibnamefont {Capasso}},\ and\ \bibinfo {author} {\bibfnamefont {Z.}~\bibnamefont {Gaburro}},\ }\bibfield  {title} {\enquote {\bibinfo {title} {Light propagation with phase discontinuities: Generalized laws of reflection and refraction},}\ }\href@noop {} {\bibfield  {journal} {\bibinfo  {journal} {Science}\ }\textbf {\bibinfo {volume} {334}},\ \bibinfo {pages} {333--337} (\bibinfo {year} {2011})}\BibitemShut {NoStop}%
\bibitem [{\citenamefont {Pfeiffer}\ and\ \citenamefont {Grbic}(2013)}]{pfeiffer2013}%
  \BibitemOpen
  \bibfield  {author} {\bibinfo {author} {\bibfnamefont {C.}~\bibnamefont {Pfeiffer}}\ and\ \bibinfo {author} {\bibfnamefont {A.}~\bibnamefont {Grbic}},\ }\bibfield  {title} {\enquote {\bibinfo {title} {Metamaterial {Huygens}' surfaces: Tailoring wave fronts with reflectionless sheets},}\ }\href@noop {} {\bibfield  {journal} {\bibinfo  {journal} {Phys. Rev. Lett.}\ }\textbf {\bibinfo {volume} {110}},\ \bibinfo {pages} {197401} (\bibinfo {year} {2013})}\BibitemShut {NoStop}%
\bibitem [{\citenamefont {Li}, \citenamefont {Singh},\ and\ \citenamefont {Sievenpiper}(2018)}]{li2018}%
  \BibitemOpen
  \bibfield  {author} {\bibinfo {author} {\bibfnamefont {A.}~\bibnamefont {Li}}, \bibinfo {author} {\bibfnamefont {S.}~\bibnamefont {Singh}},\ and\ \bibinfo {author} {\bibfnamefont {D.}~\bibnamefont {Sievenpiper}},\ }\bibfield  {title} {\enquote {\bibinfo {title} {Metasurfaces and their applications},}\ }\href@noop {} {\bibfield  {journal} {\bibinfo  {journal} {Nanophotonics}\ }\textbf {\bibinfo {volume} {7}},\ \bibinfo {pages} {989--1011} (\bibinfo {year} {2018})}\BibitemShut {NoStop}%
\bibitem [{\citenamefont {Glybovski}\ \emph {et~al.}(2016)\citenamefont {Glybovski}, \citenamefont {Tretyakov}, \citenamefont {Belov}, \citenamefont {Kivshar},\ and\ \citenamefont {Simovski}}]{glybovski2016}%
  \BibitemOpen
  \bibfield  {author} {\bibinfo {author} {\bibfnamefont {S.~B.}\ \bibnamefont {Glybovski}}, \bibinfo {author} {\bibfnamefont {S.~A.}\ \bibnamefont {Tretyakov}}, \bibinfo {author} {\bibfnamefont {P.~A.}\ \bibnamefont {Belov}}, \bibinfo {author} {\bibfnamefont {Y.~S.}\ \bibnamefont {Kivshar}},\ and\ \bibinfo {author} {\bibfnamefont {C.~R.}\ \bibnamefont {Simovski}},\ }\bibfield  {title} {\enquote {\bibinfo {title} {Metasurfaces: From microwaves to visible},}\ }\href@noop {} {\bibfield  {journal} {\bibinfo  {journal} {Phys. Rep.}\ }\textbf {\bibinfo {volume} {634}},\ \bibinfo {pages} {1--72} (\bibinfo {year} {2016})}\BibitemShut {NoStop}%
\bibitem [{\citenamefont {Cui}\ \emph {et~al.}(2014)\citenamefont {Cui}, \citenamefont {Qi}, \citenamefont {Wan}, \citenamefont {Zhao},\ and\ \citenamefont {Cheng}}]{cui2014}%
  \BibitemOpen
  \bibfield  {author} {\bibinfo {author} {\bibfnamefont {T.~J.}\ \bibnamefont {Cui}}, \bibinfo {author} {\bibfnamefont {M.~Q.}\ \bibnamefont {Qi}}, \bibinfo {author} {\bibfnamefont {X.}~\bibnamefont {Wan}}, \bibinfo {author} {\bibfnamefont {J.}~\bibnamefont {Zhao}},\ and\ \bibinfo {author} {\bibfnamefont {Q.}~\bibnamefont {Cheng}},\ }\bibfield  {title} {\enquote {\bibinfo {title} {Coding metamaterials, digital metamaterials and programmable metamaterials},}\ }\href@noop {} {\bibfield  {journal} {\bibinfo  {journal} {Light Sci. Appl.}\ }\textbf {\bibinfo {volume} {3}},\ \bibinfo {pages} {e218} (\bibinfo {year} {2014})}\BibitemShut {NoStop}%
\bibitem [{\citenamefont {Di~Renzo}\ \emph {et~al.}(2020)\citenamefont {Di~Renzo}, \citenamefont {Zappone}, \citenamefont {Debbah}, \citenamefont {Alouini}, \citenamefont {Yuen}, \citenamefont {de~Rosny},\ and\ \citenamefont {Tretyakov}}]{direnzo2020}%
  \BibitemOpen
  \bibfield  {author} {\bibinfo {author} {\bibfnamefont {M.}~\bibnamefont {Di~Renzo}}, \bibinfo {author} {\bibfnamefont {A.}~\bibnamefont {Zappone}}, \bibinfo {author} {\bibfnamefont {M.}~\bibnamefont {Debbah}}, \bibinfo {author} {\bibfnamefont {M.-S.}\ \bibnamefont {Alouini}}, \bibinfo {author} {\bibfnamefont {C.}~\bibnamefont {Yuen}}, \bibinfo {author} {\bibfnamefont {J.}~\bibnamefont {de~Rosny}},\ and\ \bibinfo {author} {\bibfnamefont {S.}~\bibnamefont {Tretyakov}},\ }\bibfield  {title} {\enquote {\bibinfo {title} {Smart radio environments empowered by reconfigurable intelligent surfaces: How it works, state of research, and the road ahead},}\ }\href@noop {} {\bibfield  {journal} {\bibinfo  {journal} {IEEE J. Sel. Areas Commun.}\ }\textbf {\bibinfo {volume} {38}},\ \bibinfo {pages} {2450--2525} (\bibinfo {year} {2020})}\BibitemShut {NoStop}%
\bibitem [{\citenamefont {Wakatsuchi}\ \emph {et~al.}(2013)\citenamefont {Wakatsuchi}, \citenamefont {Kim}, \citenamefont {Rushton},\ and\ \citenamefont {Sievenpiper}}]{wakatsuchi2013}%
  \BibitemOpen
  \bibfield  {author} {\bibinfo {author} {\bibfnamefont {H.}~\bibnamefont {Wakatsuchi}}, \bibinfo {author} {\bibfnamefont {S.}~\bibnamefont {Kim}}, \bibinfo {author} {\bibfnamefont {J.~J.}\ \bibnamefont {Rushton}},\ and\ \bibinfo {author} {\bibfnamefont {D.~F.}\ \bibnamefont {Sievenpiper}},\ }\bibfield  {title} {\enquote {\bibinfo {title} {Waveform-dependent absorbing metasurfaces},}\ }\href@noop {} {\bibfield  {journal} {\bibinfo  {journal} {Phys. Rev. Lett.}\ }\textbf {\bibinfo {volume} {111}},\ \bibinfo {pages} {245501} (\bibinfo {year} {2013})}\BibitemShut {NoStop}%
\bibitem [{\citenamefont {Ushikoshi}\ \emph {et~al.}(2023)\citenamefont {Ushikoshi}, \citenamefont {Higashiura}, \citenamefont {Tachi}, \citenamefont {Fathnan}, \citenamefont {Mahmood}, \citenamefont {Takeshita}, \citenamefont {Homma}, \citenamefont {Akram}, \citenamefont {Vellucci}, \citenamefont {Lee}, \citenamefont {Toscano}, \citenamefont {Bilotti}, \citenamefont {Christopoulos},\ and\ \citenamefont {Wakatsuchi}}]{ushikoshi2023}%
  \BibitemOpen
  \bibfield  {author} {\bibinfo {author} {\bibfnamefont {D.}~\bibnamefont {Ushikoshi}}, \bibinfo {author} {\bibfnamefont {R.}~\bibnamefont {Higashiura}}, \bibinfo {author} {\bibfnamefont {K.}~\bibnamefont {Tachi}}, \bibinfo {author} {\bibfnamefont {A.~A.}\ \bibnamefont {Fathnan}}, \bibinfo {author} {\bibfnamefont {S.}~\bibnamefont {Mahmood}}, \bibinfo {author} {\bibfnamefont {H.}~\bibnamefont {Takeshita}}, \bibinfo {author} {\bibfnamefont {H.}~\bibnamefont {Homma}}, \bibinfo {author} {\bibfnamefont {M.~R.}\ \bibnamefont {Akram}}, \bibinfo {author} {\bibfnamefont {S.}~\bibnamefont {Vellucci}}, \bibinfo {author} {\bibfnamefont {J.}~\bibnamefont {Lee}}, \bibinfo {author} {\bibfnamefont {A.}~\bibnamefont {Toscano}}, \bibinfo {author} {\bibfnamefont {F.}~\bibnamefont {Bilotti}}, \bibinfo {author} {\bibfnamefont {C.}~\bibnamefont {Christopoulos}},\ and\ \bibinfo {author} {\bibfnamefont {H.}~\bibnamefont {Wakatsuchi}},\ }\bibfield  {title} {\enquote {\bibinfo {title} {Pulse-driven self-reconfigurable
  meta-antennas},}\ }\href@noop {} {\bibfield  {journal} {\bibinfo  {journal} {Nat. Commun.}\ }\textbf {\bibinfo {volume} {14}},\ \bibinfo {pages} {633} (\bibinfo {year} {2023})}\BibitemShut {NoStop}%
\bibitem [{\citenamefont {Omori}\ \emph {et~al.}(2026)\citenamefont {Omori}, \citenamefont {Takimoto}, \citenamefont {Nagata}, \citenamefont {Fathnan}, \citenamefont {Sugiura},\ and\ \citenamefont {Wakatsuchi}}]{omori2026experimental}%
  \BibitemOpen
  \bibfield  {author} {\bibinfo {author} {\bibfnamefont {E.}~\bibnamefont {Omori}}, \bibinfo {author} {\bibfnamefont {K.}~\bibnamefont {Takimoto}}, \bibinfo {author} {\bibfnamefont {A.}~\bibnamefont {Nagata}}, \bibinfo {author} {\bibfnamefont {A.}~\bibnamefont {Fathnan}}, \bibinfo {author} {\bibfnamefont {S.}~\bibnamefont {Sugiura}},\ and\ \bibinfo {author} {\bibfnamefont {H.}~\bibnamefont {Wakatsuchi}},\ }\bibfield  {title} {\enquote {\bibinfo {title} {An experimental validation of reconfigurable intelligent surfaces achieving pulse-width-modulated singular reflection angles without external power sources},}\ }\href@noop {} {\bibfield  {journal} {\bibinfo  {journal} {IEEE Trans. Veh. Technol.}\ } (\bibinfo {year} {2026})}\BibitemShut {NoStop}%
\bibitem [{\citenamefont {Fathnan}\ \emph {et~al.}(2022)\citenamefont {Fathnan}, \citenamefont {Homma}, \citenamefont {Sugiura},\ and\ \citenamefont {Wakatsuchi}}]{fathnan2022method}%
  \BibitemOpen
  \bibfield  {author} {\bibinfo {author} {\bibfnamefont {A.~A.}\ \bibnamefont {Fathnan}}, \bibinfo {author} {\bibfnamefont {H.}~\bibnamefont {Homma}}, \bibinfo {author} {\bibfnamefont {S.}~\bibnamefont {Sugiura}},\ and\ \bibinfo {author} {\bibfnamefont {H.}~\bibnamefont {Wakatsuchi}},\ }\bibfield  {title} {\enquote {\bibinfo {title} {Method for extracting the equivalent admittance from time-varying metasurfaces and its application to self-tuned spatiotemporal wave manipulation},}\ }\href@noop {} {\bibfield  {journal} {\bibinfo  {journal} {J. Phys. D Appl. Phys.}\ }\textbf {\bibinfo {volume} {55}},\ \bibinfo {pages} {015304} (\bibinfo {year} {2022})}\BibitemShut {NoStop}%
\bibitem [{\citenamefont {Budhu}, \citenamefont {Ventresca},\ and\ \citenamefont {Grbic}(2023)}]{budhu2023}%
  \BibitemOpen
  \bibfield  {author} {\bibinfo {author} {\bibfnamefont {J.}~\bibnamefont {Budhu}}, \bibinfo {author} {\bibfnamefont {N.}~\bibnamefont {Ventresca}},\ and\ \bibinfo {author} {\bibfnamefont {A.}~\bibnamefont {Grbic}},\ }\bibfield  {title} {\enquote {\bibinfo {title} {Unit cell design for aperiodic metasurfaces},}\ }\href@noop {} {\bibfield  {journal} {\bibinfo  {journal} {IEEE Trans. Antennas Propag.}\ }\textbf {\bibinfo {volume} {71}},\ \bibinfo {pages} {7387--7394} (\bibinfo {year} {2023})}\BibitemShut {NoStop}%
\bibitem [{\citenamefont {Yang}\ \emph {et~al.}(2017)\citenamefont {Yang}, \citenamefont {Chen}, \citenamefont {Li}, \citenamefont {Zhang}, \citenamefont {Xu}, \citenamefont {Tian}, \citenamefont {Ouyang}, \citenamefont {Gu}, \citenamefont {Han},\ and\ \citenamefont {Zhang}}]{yang2017}%
  \BibitemOpen
  \bibfield  {author} {\bibinfo {author} {\bibfnamefont {Q.}~\bibnamefont {Yang}}, \bibinfo {author} {\bibfnamefont {X.}~\bibnamefont {Chen}}, \bibinfo {author} {\bibfnamefont {Y.}~\bibnamefont {Li}}, \bibinfo {author} {\bibfnamefont {X.}~\bibnamefont {Zhang}}, \bibinfo {author} {\bibfnamefont {Y.}~\bibnamefont {Xu}}, \bibinfo {author} {\bibfnamefont {Z.}~\bibnamefont {Tian}}, \bibinfo {author} {\bibfnamefont {C.}~\bibnamefont {Ouyang}}, \bibinfo {author} {\bibfnamefont {J.}~\bibnamefont {Gu}}, \bibinfo {author} {\bibfnamefont {J.}~\bibnamefont {Han}},\ and\ \bibinfo {author} {\bibfnamefont {W.}~\bibnamefont {Zhang}},\ }\bibfield  {title} {\enquote {\bibinfo {title} {Aperiodic-metamaterial-based absorber},}\ }\href@noop {} {\bibfield  {journal} {\bibinfo  {journal} {APL Mater.}\ }\textbf {\bibinfo {volume} {5}},\ \bibinfo {pages} {096107} (\bibinfo {year} {2017})}\BibitemShut {NoStop}%
\bibitem [{\citenamefont {Dal~Negro}\ and\ \citenamefont {Boriskina}(2012)}]{dalnegro2012}%
  \BibitemOpen
  \bibfield  {author} {\bibinfo {author} {\bibfnamefont {L.}~\bibnamefont {Dal~Negro}}\ and\ \bibinfo {author} {\bibfnamefont {S.~V.}\ \bibnamefont {Boriskina}},\ }\bibfield  {title} {\enquote {\bibinfo {title} {Deterministic aperiodic nanostructures for photonics and plasmonics applications},}\ }\href@noop {} {\bibfield  {journal} {\bibinfo  {journal} {Laser Photonics Rev.}\ }\textbf {\bibinfo {volume} {6}},\ \bibinfo {pages} {178--218} (\bibinfo {year} {2012})}\BibitemShut {NoStop}%
\bibitem [{\citenamefont {Jang}\ \emph {et~al.}(2018)\citenamefont {Jang}, \citenamefont {Horie}, \citenamefont {Shibukawa}, \citenamefont {Brake}, \citenamefont {Liu}, \citenamefont {Kamali}, \citenamefont {Arbabi}, \citenamefont {Ruan}, \citenamefont {Faraon},\ and\ \citenamefont {Yang}}]{jang2018}%
  \BibitemOpen
  \bibfield  {author} {\bibinfo {author} {\bibfnamefont {M.}~\bibnamefont {Jang}}, \bibinfo {author} {\bibfnamefont {Y.}~\bibnamefont {Horie}}, \bibinfo {author} {\bibfnamefont {A.}~\bibnamefont {Shibukawa}}, \bibinfo {author} {\bibfnamefont {J.}~\bibnamefont {Brake}}, \bibinfo {author} {\bibfnamefont {Y.}~\bibnamefont {Liu}}, \bibinfo {author} {\bibfnamefont {S.~M.}\ \bibnamefont {Kamali}}, \bibinfo {author} {\bibfnamefont {A.}~\bibnamefont {Arbabi}}, \bibinfo {author} {\bibfnamefont {H.}~\bibnamefont {Ruan}}, \bibinfo {author} {\bibfnamefont {A.}~\bibnamefont {Faraon}},\ and\ \bibinfo {author} {\bibfnamefont {C.}~\bibnamefont {Yang}},\ }\bibfield  {title} {\enquote {\bibinfo {title} {Wavefront shaping with disorder-engineered metasurfaces},}\ }\href@noop {} {\bibfield  {journal} {\bibinfo  {journal} {Nat. Photonics}\ }\textbf {\bibinfo {volume} {12}},\ \bibinfo {pages} {84--90} (\bibinfo {year} {2018})}\BibitemShut {NoStop}%
\bibitem [{\citenamefont {He}\ and\ \citenamefont {Chan}(2016)}]{hechan2016}%
  \BibitemOpen
  \bibfield  {author} {\bibinfo {author} {\bibfnamefont {S.}~\bibnamefont {He}}\ and\ \bibinfo {author} {\bibfnamefont {S.-H.~G.}\ \bibnamefont {Chan}},\ }\bibfield  {title} {\enquote {\bibinfo {title} {Wi-{Fi} fingerprint-based indoor positioning: Recent advances and comparisons},}\ }\href@noop {} {\bibfield  {journal} {\bibinfo  {journal} {IEEE Commun. Surveys Tuts.}\ }\textbf {\bibinfo {volume} {18}},\ \bibinfo {pages} {466--490} (\bibinfo {year} {2016})}\BibitemShut {NoStop}%
\bibitem [{\citenamefont {Wymeersch}\ \emph {et~al.}(2020)\citenamefont {Wymeersch}, \citenamefont {He}, \citenamefont {Denis}, \citenamefont {Clemente},\ and\ \citenamefont {Juntti}}]{wymeersch2020}%
  \BibitemOpen
  \bibfield  {author} {\bibinfo {author} {\bibfnamefont {H.}~\bibnamefont {Wymeersch}}, \bibinfo {author} {\bibfnamefont {J.}~\bibnamefont {He}}, \bibinfo {author} {\bibfnamefont {B.}~\bibnamefont {Denis}}, \bibinfo {author} {\bibfnamefont {A.}~\bibnamefont {Clemente}},\ and\ \bibinfo {author} {\bibfnamefont {M.}~\bibnamefont {Juntti}},\ }\bibfield  {title} {\enquote {\bibinfo {title} {Radio localization and mapping with reconfigurable intelligent surfaces: Challenges, opportunities, and research directions},}\ }\href@noop {} {\bibfield  {journal} {\bibinfo  {journal} {IEEE Veh. Technol. Mag.}\ }\textbf {\bibinfo {volume} {15}},\ \bibinfo {pages} {52--61} (\bibinfo {year} {2020})}\BibitemShut {NoStop}%
\bibitem [{\citenamefont {Preradovic}\ and\ \citenamefont {Karmakar}(2010)}]{preradovic2010}%
  \BibitemOpen
  \bibfield  {author} {\bibinfo {author} {\bibfnamefont {S.}~\bibnamefont {Preradovic}}\ and\ \bibinfo {author} {\bibfnamefont {N.~C.}\ \bibnamefont {Karmakar}},\ }\bibfield  {title} {\enquote {\bibinfo {title} {Chipless {RFID}: Bar code of the future},}\ }\href@noop {} {\bibfield  {journal} {\bibinfo  {journal} {IEEE Microw. Mag.}\ }\textbf {\bibinfo {volume} {11}},\ \bibinfo {pages} {87--97} (\bibinfo {year} {2010})}\BibitemShut {NoStop}%
\bibitem [{\citenamefont {del Hougne}\ \emph {et~al.}(2018)\citenamefont {del Hougne}, \citenamefont {Imani}, \citenamefont {Fink}, \citenamefont {Smith},\ and\ \citenamefont {Lerosey}}]{delhougne2018}%
  \BibitemOpen
  \bibfield  {author} {\bibinfo {author} {\bibfnamefont {P.}~\bibnamefont {del Hougne}}, \bibinfo {author} {\bibfnamefont {M.~F.}\ \bibnamefont {Imani}}, \bibinfo {author} {\bibfnamefont {M.}~\bibnamefont {Fink}}, \bibinfo {author} {\bibfnamefont {D.~R.}\ \bibnamefont {Smith}},\ and\ \bibinfo {author} {\bibfnamefont {G.}~\bibnamefont {Lerosey}},\ }\bibfield  {title} {\enquote {\bibinfo {title} {Precise localization of multiple noncooperative objects in a disordered cavity by wave front shaping},}\ }\href@noop {} {\bibfield  {journal} {\bibinfo  {journal} {Phys. Rev. Lett.}\ }\textbf {\bibinfo {volume} {121}},\ \bibinfo {pages} {063901} (\bibinfo {year} {2018})}\BibitemShut {NoStop}%
\bibitem [{\citenamefont {Liu}\ \emph {et~al.}(2022)\citenamefont {Liu}, \citenamefont {Cui}, \citenamefont {Masouros}, \citenamefont {Xu}, \citenamefont {Han}, \citenamefont {Eldar},\ and\ \citenamefont {Buzzi}}]{liu2022}%
  \BibitemOpen
  \bibfield  {author} {\bibinfo {author} {\bibfnamefont {F.}~\bibnamefont {Liu}}, \bibinfo {author} {\bibfnamefont {Y.}~\bibnamefont {Cui}}, \bibinfo {author} {\bibfnamefont {C.}~\bibnamefont {Masouros}}, \bibinfo {author} {\bibfnamefont {J.}~\bibnamefont {Xu}}, \bibinfo {author} {\bibfnamefont {T.~X.}\ \bibnamefont {Han}}, \bibinfo {author} {\bibfnamefont {Y.~C.}\ \bibnamefont {Eldar}},\ and\ \bibinfo {author} {\bibfnamefont {S.}~\bibnamefont {Buzzi}},\ }\bibfield  {title} {\enquote {\bibinfo {title} {Integrated sensing and communications: Toward dual-functional wireless networks for {6G} and beyond},}\ }\href@noop {} {\bibfield  {journal} {\bibinfo  {journal} {IEEE J. Sel. Areas Commun.}\ }\textbf {\bibinfo {volume} {40}},\ \bibinfo {pages} {1728--1767} (\bibinfo {year} {2022})}\BibitemShut {NoStop}%
\bibitem [{\citenamefont {Smith}\ \emph {et~al.}(2024{\natexlab{a}})\citenamefont {Smith}, \citenamefont {Myers}, \citenamefont {Kaplan},\ and\ \citenamefont {Goodman-Strauss}}]{smith2024hat}%
  \BibitemOpen
  \bibfield  {author} {\bibinfo {author} {\bibfnamefont {D.}~\bibnamefont {Smith}}, \bibinfo {author} {\bibfnamefont {J.~S.}\ \bibnamefont {Myers}}, \bibinfo {author} {\bibfnamefont {C.~S.}\ \bibnamefont {Kaplan}},\ and\ \bibinfo {author} {\bibfnamefont {C.}~\bibnamefont {Goodman-Strauss}},\ }\bibfield  {title} {\enquote {\bibinfo {title} {An aperiodic monotile},}\ }\href@noop {} {\bibfield  {journal} {\bibinfo  {journal} {Comb. Theory}\ }\textbf {\bibinfo {volume} {4}},\ \bibinfo {pages} {6} (\bibinfo {year} {2024}{\natexlab{a}})}\BibitemShut {NoStop}%
\bibitem [{\citenamefont {Smith}\ \emph {et~al.}(2024{\natexlab{b}})\citenamefont {Smith}, \citenamefont {Myers}, \citenamefont {Kaplan},\ and\ \citenamefont {Goodman-Strauss}}]{smith2024spectre}%
  \BibitemOpen
  \bibfield  {author} {\bibinfo {author} {\bibfnamefont {D.}~\bibnamefont {Smith}}, \bibinfo {author} {\bibfnamefont {J.~S.}\ \bibnamefont {Myers}}, \bibinfo {author} {\bibfnamefont {C.~S.}\ \bibnamefont {Kaplan}},\ and\ \bibinfo {author} {\bibfnamefont {C.}~\bibnamefont {Goodman-Strauss}},\ }\bibfield  {title} {\enquote {\bibinfo {title} {A chiral aperiodic monotile},}\ }\href@noop {} {\bibfield  {journal} {\bibinfo  {journal} {Comb. Theory}\ }\textbf {\bibinfo {volume} {4}},\ \bibinfo {pages} {13} (\bibinfo {year} {2024}{\natexlab{b}})}\BibitemShut {NoStop}%
\bibitem [{\citenamefont {Penrose}(1974)}]{penrose1974}%
  \BibitemOpen
  \bibfield  {author} {\bibinfo {author} {\bibfnamefont {R.}~\bibnamefont {Penrose}},\ }\bibfield  {title} {\enquote {\bibinfo {title} {The role of aesthetics in pure and applied mathematical research},}\ }\href@noop {} {\bibfield  {journal} {\bibinfo  {journal} {Bull. Inst. Math. Appl.}\ }\textbf {\bibinfo {volume} {10}},\ \bibinfo {pages} {266--271} (\bibinfo {year} {1974})}\BibitemShut {NoStop}%
\bibitem [{\citenamefont {Vardeny}, \citenamefont {Nahata},\ and\ \citenamefont {Agrawal}(2013)}]{vardeny2013}%
  \BibitemOpen
  \bibfield  {author} {\bibinfo {author} {\bibfnamefont {Z.~V.}\ \bibnamefont {Vardeny}}, \bibinfo {author} {\bibfnamefont {A.}~\bibnamefont {Nahata}},\ and\ \bibinfo {author} {\bibfnamefont {A.}~\bibnamefont {Agrawal}},\ }\bibfield  {title} {\enquote {\bibinfo {title} {Optics of photonic quasicrystals},}\ }\href@noop {} {\bibfield  {journal} {\bibinfo  {journal} {Nat. Photonics}\ }\textbf {\bibinfo {volume} {7}},\ \bibinfo {pages} {177--187} (\bibinfo {year} {2013})}\BibitemShut {NoStop}%
\bibitem [{\citenamefont {Socolar}(2023)}]{socolar2023}%
  \BibitemOpen
  \bibfield  {author} {\bibinfo {author} {\bibfnamefont {J.~E.~S.}\ \bibnamefont {Socolar}},\ }\bibfield  {title} {\enquote {\bibinfo {title} {Quasicrystalline structure of the hat monotile tilings},}\ }\href@noop {} {\bibfield  {journal} {\bibinfo  {journal} {Phys. Rev. B}\ }\textbf {\bibinfo {volume} {108}},\ \bibinfo {pages} {224109} (\bibinfo {year} {2023})}\BibitemShut {NoStop}%
\bibitem [{\citenamefont {Schirmann}\ \emph {et~al.}(2024)\citenamefont {Schirmann}, \citenamefont {Franca}, \citenamefont {Flicker},\ and\ \citenamefont {Grushin}}]{schirmann2024}%
  \BibitemOpen
  \bibfield  {author} {\bibinfo {author} {\bibfnamefont {J.}~\bibnamefont {Schirmann}}, \bibinfo {author} {\bibfnamefont {S.}~\bibnamefont {Franca}}, \bibinfo {author} {\bibfnamefont {F.}~\bibnamefont {Flicker}},\ and\ \bibinfo {author} {\bibfnamefont {A.~G.}\ \bibnamefont {Grushin}},\ }\bibfield  {title} {\enquote {\bibinfo {title} {Physical properties of an aperiodic monotile with graphene-like features, chirality, and zero modes},}\ }\href@noop {} {\bibfield  {journal} {\bibinfo  {journal} {Phys. Rev. Lett.}\ }\textbf {\bibinfo {volume} {132}},\ \bibinfo {pages} {086402} (\bibinfo {year} {2024})}\BibitemShut {NoStop}%
\bibitem [{\citenamefont {Moritake}\ \emph {et~al.}(2026)\citenamefont {Moritake}, \citenamefont {Takiguchi}, \citenamefont {Aihara},\ and\ \citenamefont {Notomi}}]{moritake2026chiral}%
  \BibitemOpen
  \bibfield  {author} {\bibinfo {author} {\bibfnamefont {Y.}~\bibnamefont {Moritake}}, \bibinfo {author} {\bibfnamefont {M.}~\bibnamefont {Takiguchi}}, \bibinfo {author} {\bibfnamefont {T.}~\bibnamefont {Aihara}},\ and\ \bibinfo {author} {\bibfnamefont {M.}~\bibnamefont {Notomi}},\ }\bibfield  {title} {\enquote {\bibinfo {title} {Chiral diffraction from aperiodic monotile structure},}\ }\href {https://doi.org/10.1038/s41467-026-75023-7} {\bibfield  {journal} {\bibinfo  {journal} {Nat. Commun.}\ }\textbf {\bibinfo {volume} {17}},\ \bibinfo {pages} {6085} (\bibinfo {year} {2026})}\BibitemShut {NoStop}%
\bibitem [{\citenamefont {Qiang}\ \emph {et~al.}(2025)\citenamefont {Qiang}, \citenamefont {Fang}, \citenamefont {Wu}, \citenamefont {Chen},\ and\ \citenamefont {Wang}}]{qiang2025}%
  \BibitemOpen
  \bibfield  {author} {\bibinfo {author} {\bibfnamefont {Y.}~\bibnamefont {Qiang}}, \bibinfo {author} {\bibfnamefont {X.}~\bibnamefont {Fang}}, \bibinfo {author} {\bibfnamefont {R.-X.}\ \bibnamefont {Wu}}, \bibinfo {author} {\bibfnamefont {Q.}~\bibnamefont {Chen}},\ and\ \bibinfo {author} {\bibfnamefont {W.}~\bibnamefont {Wang}},\ }\bibfield  {title} {\enquote {\bibinfo {title} {Application of aperiodic `{Einstein}' monotile in phased arrays with limited beam scanning range},}\ }\href@noop {} {\bibfield  {journal} {\bibinfo  {journal} {IEEE Open J. Antennas Propag.}\ }\textbf {\bibinfo {volume} {6}},\ \bibinfo {pages} {283--292} (\bibinfo {year} {2025})}\BibitemShut {NoStop}%
\bibitem [{\citenamefont {Lee}, \citenamefont {Jeon},\ and\ \citenamefont {Shin}(2025)}]{lee2025aperiodic}%
  \BibitemOpen
  \bibfield  {author} {\bibinfo {author} {\bibfnamefont {M.}~\bibnamefont {Lee}}, \bibinfo {author} {\bibfnamefont {S.}~\bibnamefont {Jeon}},\ and\ \bibinfo {author} {\bibfnamefont {J.}~\bibnamefont {Shin}},\ }\bibfield  {title} {\enquote {\bibinfo {title} {Angle- and polarization-tolerant metasurface designs based on the aperiodic tiling},}\ }\href {https://doi.org/10.1002/adom.202500455} {\bibfield  {journal} {\bibinfo  {journal} {Adv. Opt. Mater.}\ }\textbf {\bibinfo {volume} {13}},\ \bibinfo {pages} {2500455} (\bibinfo {year} {2025})}\BibitemShut {NoStop}%
\bibitem [{\citenamefont {Donoho}(2006)}]{donoho2006}%
  \BibitemOpen
  \bibfield  {author} {\bibinfo {author} {\bibfnamefont {D.~L.}\ \bibnamefont {Donoho}},\ }\bibfield  {title} {\enquote {\bibinfo {title} {Compressed sensing},}\ }\href@noop {} {\bibfield  {journal} {\bibinfo  {journal} {IEEE Trans. Inf. Theory}\ }\textbf {\bibinfo {volume} {52}},\ \bibinfo {pages} {1289--1306} (\bibinfo {year} {2006})}\BibitemShut {NoStop}%
\bibitem [{\citenamefont {Cand{\`e}s}\ and\ \citenamefont {Wakin}(2008)}]{candes2008}%
  \BibitemOpen
  \bibfield  {author} {\bibinfo {author} {\bibfnamefont {E.~J.}\ \bibnamefont {Cand{\`e}s}}\ and\ \bibinfo {author} {\bibfnamefont {M.~B.}\ \bibnamefont {Wakin}},\ }\bibfield  {title} {\enquote {\bibinfo {title} {An introduction to compressive sampling},}\ }\href@noop {} {\bibfield  {journal} {\bibinfo  {journal} {IEEE Signal Process. Mag.}\ }\textbf {\bibinfo {volume} {25}},\ \bibinfo {pages} {21--30} (\bibinfo {year} {2008})}\BibitemShut {NoStop}%
\bibitem [{\citenamefont {Vershynin}(2018)}]{vershynin2018}%
  \BibitemOpen
  \bibfield  {author} {\bibinfo {author} {\bibfnamefont {R.}~\bibnamefont {Vershynin}},\ }\href@noop {} {\emph {\bibinfo {title} {High-Dimensional Probability: An Introduction with Applications in Data Science}}}\ (\bibinfo  {publisher} {Cambridge University Press},\ \bibinfo {address} {Cambridge},\ \bibinfo {year} {2018})\BibitemShut {NoStop}%
\bibitem [{\citenamefont {Li}\ \emph {et~al.}(2017)\citenamefont {Li}, \citenamefont {Luo}, \citenamefont {Wakatsuchi}, \citenamefont {Kim},\ and\ \citenamefont {Sievenpiper}}]{li2017nonlinear}%
  \BibitemOpen
  \bibfield  {author} {\bibinfo {author} {\bibfnamefont {A.}~\bibnamefont {Li}}, \bibinfo {author} {\bibfnamefont {Z.}~\bibnamefont {Luo}}, \bibinfo {author} {\bibfnamefont {H.}~\bibnamefont {Wakatsuchi}}, \bibinfo {author} {\bibfnamefont {S.}~\bibnamefont {Kim}},\ and\ \bibinfo {author} {\bibfnamefont {D.~F.}\ \bibnamefont {Sievenpiper}},\ }\bibfield  {title} {\enquote {\bibinfo {title} {Nonlinear, active, and tunable metasurfaces for advanced electromagnetics applications},}\ }\href@noop {} {\bibfield  {journal} {\bibinfo  {journal} {IEEE Access}\ }\textbf {\bibinfo {volume} {5}},\ \bibinfo {pages} {27439--27452} (\bibinfo {year} {2017})}\BibitemShut {NoStop}%
\bibitem [{\citenamefont {Zharov}, \citenamefont {Shadrivov},\ and\ \citenamefont {Kivshar}(2003)}]{kivshar2003nonlinear}%
  \BibitemOpen
  \bibfield  {author} {\bibinfo {author} {\bibfnamefont {A.~A.}\ \bibnamefont {Zharov}}, \bibinfo {author} {\bibfnamefont {I.~V.}\ \bibnamefont {Shadrivov}},\ and\ \bibinfo {author} {\bibfnamefont {Y.~S.}\ \bibnamefont {Kivshar}},\ }\bibfield  {title} {\enquote {\bibinfo {title} {Nonlinear properties of left-handed metamaterials},}\ }\href@noop {} {\bibfield  {journal} {\bibinfo  {journal} {Phys. Rev. Lett.}\ }\textbf {\bibinfo {volume} {91}},\ \bibinfo {pages} {037401} (\bibinfo {year} {2003})}\BibitemShut {NoStop}%
\bibitem [{\citenamefont {Vellucci}\ \emph {et~al.}(2020)\citenamefont {Vellucci}, \citenamefont {Monti}, \citenamefont {Barbuto}, \citenamefont {Toscano},\ and\ \citenamefont {Bilotti}}]{vellucci2019waveform}%
  \BibitemOpen
  \bibfield  {author} {\bibinfo {author} {\bibfnamefont {S.}~\bibnamefont {Vellucci}}, \bibinfo {author} {\bibfnamefont {A.}~\bibnamefont {Monti}}, \bibinfo {author} {\bibfnamefont {M.}~\bibnamefont {Barbuto}}, \bibinfo {author} {\bibfnamefont {A.}~\bibnamefont {Toscano}},\ and\ \bibinfo {author} {\bibfnamefont {F.}~\bibnamefont {Bilotti}},\ }\bibfield  {title} {\enquote {\bibinfo {title} {Waveform-selective mantle cloaks for intelligent antennas},}\ }\href@noop {} {\bibfield  {journal} {\bibinfo  {journal} {IEEE Trans. Antennas Propag.}\ }\textbf {\bibinfo {volume} {68}},\ \bibinfo {pages} {1717--1725} (\bibinfo {year} {2020})}\BibitemShut {NoStop}%
\bibitem [{\citenamefont {Barbuto}\ \emph {et~al.}(2022)\citenamefont {Barbuto}, \citenamefont {Hamzavi-Zarghani}, \citenamefont {Longhi}, \citenamefont {Monti}, \citenamefont {Ramaccia}, \citenamefont {Vellucci}, \citenamefont {Toscano},\ and\ \citenamefont {Bilotti}}]{barbuto2021metasurfaces}%
  \BibitemOpen
  \bibfield  {author} {\bibinfo {author} {\bibfnamefont {M.}~\bibnamefont {Barbuto}}, \bibinfo {author} {\bibfnamefont {Z.}~\bibnamefont {Hamzavi-Zarghani}}, \bibinfo {author} {\bibfnamefont {M.}~\bibnamefont {Longhi}}, \bibinfo {author} {\bibfnamefont {A.}~\bibnamefont {Monti}}, \bibinfo {author} {\bibfnamefont {D.}~\bibnamefont {Ramaccia}}, \bibinfo {author} {\bibfnamefont {S.}~\bibnamefont {Vellucci}}, \bibinfo {author} {\bibfnamefont {A.}~\bibnamefont {Toscano}},\ and\ \bibinfo {author} {\bibfnamefont {F.}~\bibnamefont {Bilotti}},\ }\bibfield  {title} {\enquote {\bibinfo {title} {Metasurfaces 3.0: a new paradigm for enabling smart electromagnetic environments},}\ }\href@noop {} {\bibfield  {journal} {\bibinfo  {journal} {IEEE Trans. Antennas Propag.}\ }\textbf {\bibinfo {volume} {70}},\ \bibinfo {pages} {8883--8897} (\bibinfo {year} {2022})}\BibitemShut {NoStop}%
\bibitem [{\citenamefont {Takeshita}\ \emph {et~al.}(2024)\citenamefont {Takeshita}, \citenamefont {Fathnan}, \citenamefont {Nita}, \citenamefont {Nagata}, \citenamefont {Sugiura},\ and\ \citenamefont {Wakatsuchi}}]{takeshita2024frequency}%
  \BibitemOpen
  \bibfield  {author} {\bibinfo {author} {\bibfnamefont {H.}~\bibnamefont {Takeshita}}, \bibinfo {author} {\bibfnamefont {A.~A.}\ \bibnamefont {Fathnan}}, \bibinfo {author} {\bibfnamefont {D.}~\bibnamefont {Nita}}, \bibinfo {author} {\bibfnamefont {A.}~\bibnamefont {Nagata}}, \bibinfo {author} {\bibfnamefont {S.}~\bibnamefont {Sugiura}},\ and\ \bibinfo {author} {\bibfnamefont {H.}~\bibnamefont {Wakatsuchi}},\ }\bibfield  {title} {\enquote {\bibinfo {title} {Frequency-hopping wave engineering with metasurfaces},}\ }\href@noop {} {\bibfield  {journal} {\bibinfo  {journal} {Nat. Commun.}\ }\textbf {\bibinfo {volume} {15}},\ \bibinfo {pages} {196} (\bibinfo {year} {2024})}\BibitemShut {NoStop}%
\end{thebibliography}

%

\end{document}